\documentclass[conference]{IEEEtran}

\usepackage{amsmath,amssymb,amsfonts}
\usepackage{algorithmic}
\usepackage{graphicx}
\usepackage{textcomp}
\usepackage{xcolor}
\usepackage{microtype}
\usepackage{hyperref}
\usepackage[capitalise]{cleveref}
\usepackage{listings}
\usepackage{booktabs}
\usepackage{tabularx}
\usepackage{adjustbox}
\usepackage[british]{babel}
\usepackage{csquotes}
\usepackage[nolist]{acronym}
\usepackage{dblfloatfix}
\usepackage{balance}
\usepackage{makecell}

\usepackage{cite}

\definecolor{mygray}{rgb}{0.4,0.4,0.4}
\definecolor{mygreen}{rgb}{0,0.8,0.6}
\definecolor{myorange}{rgb}{1.0,0.4,0}

\newcolumntype{Y}{>{\centering\arraybackslash}X}
\newcolumntype{R}{>{\raggedleft\arraybackslash}X}

\begin{acronym}[PWC]
	\acro{pwc}[PWC]{pointwise convolution}
	\acro{fc}[FC]{fully-connected}
	\acro{dsc}[DSConv]{depthwise-separable convolution}
	\acro{simd}[SIMD]{Single Instruction, Multiple Data}
	\acro{nn}[NN]{neural network}
	\acro{dscnn}[DS-CNN]{depthwise separable convolutional neural network}
	\acro{csr}[CSR]{Compressed Storage Row}
	\acro{bcsr}[BCSR]{Block Compressed Storage Row}
	\acro{ri}[RI]{Relative Indexing}
	\acro{dbe}[DBE]{Dynamic Bitwidth Extension}
	\acro{dle}[DLE]{Delta-Linear Encoding}
	\acro{ib}[IB]{Index Buffering}
	\acro{vb}[VB]{Value Buffering}
	\acro{ds}[ds]{depthwise separable}
	\acro{tcm}[TCM]{tightly-coupled memory}
	\acro{dcsr}[dCSR]{Delta-Compressed Storage Row}
	\acro{mve}[MVE]{M-Profile Vector Extension}
	\acro{isa}[ISA]{Instruction Set Architecture}
	\acro{pmu}[PMU]{Performance Monitoring Unit}
	\acro{npu}[NPU]{Neural Processing Unit}
	\acro{spmv}[SpMV]{sparse matrix-vector multiplication}
	\acro{spmm}[SpMM]{sparse matrix-matrix multiplication}
	\acro{tflite}[TFlite]{Tensorflow lite}
	\acro{bram}[BRAM]{Block RAM}
	\acro{s}[S]{small}
	\acro{m}[M]{medium}
	\acro{l}[L]{large}
	\acro{bl}[BL]{baseline}
	\acro{dsp}[DSP]{Digital Signal Processing}
	\acro{mac}[MAC]{Multiply-Accumulate}
\end{acronym}

\def\BibTeX{{\rm B\kern-.05em{\sc i\kern-.025em b}\kern-.08em
    T\kern-.1667em\lower.7ex\hbox{E}\kern-.125emX}}

\begin{document}
\bstctlcite{IEEEexample:BSTcontrol}

\title{dCSR: A Memory-Efficient Sparse Matrix Representation for Parallel Neural Network Inference}

\iftrue
{
\author{\IEEEauthorblockN{1\textsuperscript{st} Elias Trommer}
	\IEEEauthorblockA{\textit{Infineon Technologies} \\
		\textit{TU Dresden}\\
		Dresden, Germany \\
		elias.trommer@infineon.com}
	\and
	\IEEEauthorblockN{2\textsuperscript{nd} Bernd Waschneck}
	\IEEEauthorblockA{\textit{Infineon Technologies} \\
		Dresden, Germany \\
		bernd.waschneck@infineon.com}
	\and
	\IEEEauthorblockN{3\textsuperscript{rd} Akash Kumar}
	\IEEEauthorblockA{\textit{Center for advancing electronics (cfaed)} \\
		\textit{TU Dresden}\\
		Dresden, Germany \\
		akash.kumar@tu-dresden.de}
}
}
\else
{
\author{\IEEEauthorblockN{1\textsuperscript{st} Anonymous Author(s)}
	\IEEEauthorblockA{\textit{Affiliation} \\
		Address \\
		email}
}
}
\fi

\maketitle

\begin{abstract}
Reducing the memory footprint of \aclp{nn} is a crucial prerequisite for deploying them in small and low-cost embedded devices. Network parameters can often be reduced significantly through pruning. We discuss how to best represent the indexing overhead of sparse networks for the coming generation of \ac{simd}-capable microcontrollers. From this, we develop \ac{dcsr}, a storage format for sparse matrices that allows for both low overhead storage and fast inference on embedded systems with wide \ac{simd} units. We demonstrate our method on an ARM Cortex-M55 MCU prototype with \ac{mve}. A comparison of memory consumption and throughput shows that our method achieves competitive compression ratios and increases throughput over dense methods by up to $2.9 \times$ for \ac{spmv}-based kernels and $1.06 \times$ for \ac{spmm}. This is accomplished through handling the generation of index information directly in the \ac{simd} unit, leading to an increase in effective memory bandwidth.
\end{abstract}
\acresetall 
\begin{IEEEkeywords}
pruning, sparse neural networks, simd, embedded systems, compression
\end{IEEEkeywords}

\section{Introduction}
With \ac{nn} applications moving closer and closer to the edge, ever smaller devices are running them. Use cases like wakeword detection~\cite{chen2014kws} or the processing of sensor data~\cite{meng2020survey} have sparked a growing interest in very small \acp{nn}, run on deeply embedded systems, often as one task among others. On these devices, memory consumption is a major concern. Memory consumption drives implementation cost through chip area that has to be dedicated to it. It also strongly impacts power consumption, since it is memory transactions---rather than compute operations---that drive the power demand of an embedded application~\cite{horowitz2014energy}. These reasons make the memory usage of models a major concern when deploying \acp{nn} on embedded systems. At the same time the increasing availability of advanced \ac{simd} capabilities on small devices~\cite{skillman2020m55, riscvv} allows for much faster, parallel computation and widens the memory-performance gap~\cite{patterson1997} along with the disparity between overall available compute and memory capacity in the system.\par
While a reduction of model parameters can often be achieved through architectures that are highly specialized for a certain task~\cite{wong2020tinyspeech}, pruning of weights in existing models might offer a more accessible alternative for reducing parameter counts. Different from dedicated architectures, pruning is applicable to a wide range of models without expert knowledge and orthogonal to other common techniques like quantization~\cite{tung2018clipq}. The use of unstructured pruning, however, is often discouraged because the induced sparsity cannot easily be exploited for gains in memory consumption or throughput. A storage scheme that aims to eliminate the zero values in the pruned network requires additional indexing information to regenerate the position of each non-zero element in a layer's weight matrix. This added overhead can easily outweigh the underlying data and impose a significant memory penalty for pruned networks~\cite{zhu2017prune}. While improving the throughput of sparse matrix calculations has long been an area of active research~\cite{liu2015csr5}, there is less work on reducing their memory consumption. Many of the techniques developed for computation on sparse networks target desktop-class processors or hardware accelerators. To the best of our knowledge, there are no solutions that target embedded processors with parallel \acp{isa}. We find, however, that quantized networks and embedded \ac{simd} units are so vastly different from their bigger siblings that the existing body of work often does not carry over to them. Sparse parallel computation on small devices is an important building block for inference on the edge; it does, however, carry its own, unique set of challenges which merit a new approach to sparse matrix storage.\par
In this work, we therefore propose \emph{\ac{dcsr}}. With \ac{dcsr}, we create a lightweight solution that enables low-overhead storage of sparse \ac{nn} layers with minimal effort required for extracting the sparse index structure. To achieve this, \ac{dcsr} utilizes embedded \ac{simd} instructions in order to parallelize index generation as well as sparse inference. Different to prior work~\cite{parashar2017, zhang2016CambriconX, han2016eie}, \ac{dcsr} is a pure software implementation and does not rely on additional hardware, significantly lowering the effort required to implement it in a real-world application. Similar to many compression schemes, groups of index elements in \ac{dcsr} can be of varying bitwidth. \ac{dcsr} differs in that it makes sure that each part of the representation is byte-aligned in memory.\par
Our contributions are:
\begin{itemize}
	\item A detailed analysis of the distinct features of embedded hardware for \ac{nn} inference and the impact of pruning on a set of \ac{nn} architectures when deployed.
	\item A matrix encoding that enables pruning as a means of lowering network parameter counts. We reduce \ac{nn} size by up to $74\%$ while at the same time yielding a speedup of up to $2.96$ over dense inference for certain layer types in our evaluation. This is achieved through a unique balance of a low memory footprint and fast extraction.
\end{itemize}

\section{Background}

\subsection{Compressed Storage Row}
\begin{figure}[tb]
	\includegraphics[width=\linewidth]{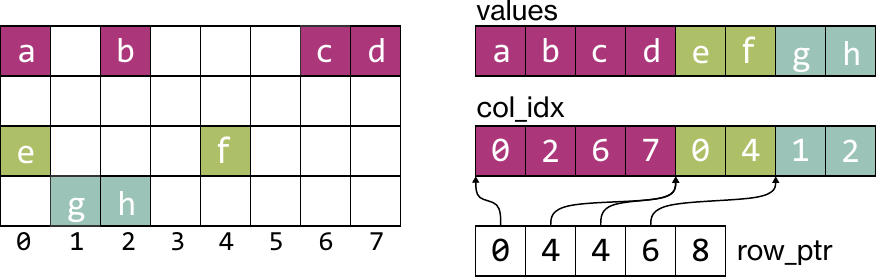}
	\caption{Conversion of a sparse matrix to CSR}
	\label{fig:csr}
\end{figure}

The most widely used format to encode sparse matrices is the \ac{csr} format. \ac{csr} uses three arrays to represent a sparse matrix as shown in~\cref{fig:csr}: the \texttt{values} array contains the non-zero values of the matrix in a contiguous array. The \texttt{col\_idx} array holds the column index of each non-zero value. Lastly, the \texttt{row\_ptr} array contains the start index of each row in the \texttt{values} and \texttt{col\_idx} array. Note that the per-element size in the column index array is determined by the number of columns, while the magnitude of the row pointers is upper-bounded by the number of non-zero values. A variant of \ac{csr} is \ac{bcsr}, in which non-zero blocks of fixed size are stored instead of single values. This increases storage overhead because not all zero elements can be eliminated. The format is sometimes preferred because it can speed up calculations, particularly for matrices with an inherent structure~\cite{williams2007}.

\subsection{Relative Indexing}
A method to reduce the overhead of \ac{csr} and its derivatives that is commonly seen in sparse network accelerators is the encoding of the distance between two adjacent non-zero elements, rather than their absolute column indices~\cite{han2016deep, parashar2017}. Particularly for the sparsities seen in \ac{nn} weight matrices, which are typically much lower than in scientific sparse matrix applications, Relative Indexing yields a significant saving in memory. In this representation, each column index difference is represented in a fixed number of bits. If there are gaps that are too large to be encoded in this fixed size, a padding value of zero is inserted in them.

\subsection{General-Purpose Compression}
Another way of reducing the memory footprint of sparse networks that sees some use in practical scenarios is simply applying a generic data compression algorithm after pruning when deploying it over a low-bandwidth channel. Because of the large number of zeroes, the sparse network can achieve higher compression ratios than its dense counterpart. After transmission, the network is extracted on the device, and inference is run the same as on a dense network. Deflate~\cite{deflate1996}, the most widely used general-purpose data compression algorithm consists of two stages. The LZSS stage initially identifies repeated sequences of symbols and replaces each repetition with a backward reference to the first occurrence of the sequence. The Huffman coding stage then assigns symbols of different bitwidth to symbols in the uncompressed input, depending on their frequency. In \cite{han2016deep}, Huffman coding is used to further increase the compression ratio of a sparse network.
\section{Methods}
Relative Indexing already targets encoding of sparse matrices for memory-constrained systems. It is a robust and simple method that can achieve good compression ratios. However, we identify two shortcomings that make it hard to use outside of dedicated hardware.

\paragraph{Inter-element dependency} To calculate the column index of an element, the cumulative sum of all previous relative indices in the same row needs to be known. While this is not a problem in purpose-built hardware~\cite{han2016eie, zhang2016CambriconX} and for floating-point inference~\cite{kourtis2008}, it becomes an issue for quantized inference on embedded systems. To understand why that is, we need to look at one of the peculiarities of quantized networks: the different sizes of accumulator register and weights/activation values. Different sizes are required because products of weights and activations can be expected to overflow the bounds of their base type. This seemingly small detail has a large impact on the iteration order: for quantized inference, iteration needs to be done \emph{horizontally}, rather than \emph{vertically}. Because \ac{simd} units for quantized operation accumulate over several multiplication results at once~\cite{gautschi2017}, they process adjacent weight-activation pairs \emph{within one row} in parallel. In contrast, GPUs and desktop CPUs typically process multiple \emph{adjacent rows} in parallel, one column at a time. This makes the ability to recover the indices of adjacent elements in one row in parallel highly important for good utilization of parallel hardware in an embedded context.

\paragraph{Computational Overhead} For higher sparsities, the amount of padding inserted by the Relative Indexing Algorithm will grow. Even when assuming a perfect decoding stage, this directly translates into degraded throughput because each padding element also causes a multiplication with zero and a memory access into the activations during inference.\par

Other idiosyncrasies of quantized inference on embedded \ac{simd} units that make a large number of common optimization techniques developed for desktop applications intractable are the re-quantization from 32-bit to 8-bit of values after inference and the format of scatter/gather instructions for 8-bit elements. For ARM \ac{mve}, currently the most mature embedded \ac{simd} \ac{isa} (the RISC-V Vector Extension still has ``draft'' status at the time of writing), these instructions only handle unsigned 8-bit offset values. The first means that the cost for well-established performance optimizations like loop tiling and every other technique that relies on the output array for intermediate storage~\cite{elsen2020fast} would be vastly increased through repeated quantizations and re-quantizations of intermediate results. The second implies that storing absolute indices of array elements like it is done in \ac{csr} is detrimental to performance if the number of columns may exceed 256. Indices need to be decomposed into a per-element offset that must not exceed the range $\left[0, 255\right]$ and an adjusted array base pointer that is passed to the scatter/gather instruction. \ac{csr} would require such a decomposition at runtime for every index element, creating significant overhead for index generation.\par

We address these points with a combination of two different techniques. One is \ac{dle}, an encoding scheme that ensures that per-element values are kept small for reduced storage overhead and independent of each other for parallel computation. The second method is \ac{dbe}, a decomposition of elements that allows for per-element values to be of varying bitwidth and still retain proper memory alignment.
\subsection{Delta-Linear Encoding}
While the row index array in \ac{csr} only grows with $\mathcal{O}(n)$ of the matrix dimensions, the column indices array grows quadratically. In practice, this means that the main emphasis of optimizing for low memory consumption should be put on reducing the size of the column indices array. As stated above, there must not be a dependency between adjacent elements in the index calculation. In order to achieve this, we first \emph{predict} an elements index using a linear mapping of its position in the column indices array and only store the deviation (the \emph{delta} value) from that prediction.\par
Assume a column index $c_i$ that is stored at position $i$ in the column indices array. We decompose $c_i$ into a mapping function $f(i)$ and the deviation from it: $\Delta c_i =  c_i - f(i)$. While sparse matrices with an inherent structure might require different choices for $f(i)$, we find that for the low-magnitude pruning scheme used, weights do not seem to be concentrated in certain areas of the weight matrix. Instead, they exhibit a uniformly random pattern. This makes a linear mapping $f(i) = i \cdot m + n$ a sensible default. We can derive the slope of this linear function as the average distance between two adjacent elements from the number of non-zero elements in the row of the sparse matrix $k_s$ and the row length of the dense matrix $k_d$. This gives us a slope of $m = \left\lfloor k_d/k_s \right\rceil$ that will produce relatively small deviations of values for $\Delta c_i$ within one row. The slope is rounded to the nearest integer to enable integer-only runtime calculations.\par
For practical applications, we group several consecutive delta elements into \ac{simd} runs $S_j$ of length $g$. This gives us the group-wise linear mapping function
\begin{equation}
c_i = \left\lfloor\frac{k_d}{k_s}\right\rceil \cdot \left(i \bmod g \right) + \Delta c_i
\end{equation}
In this representation, only the $\Delta c_i$ part contains information that needs to be stored explicitly, while the first part of the equation can be inferred at runtime. For each \ac{simd} group $S_j = \lbrace \Delta c_i, \Delta c_{i+1}, \dots, \Delta c_{i+g-1}\rbrace$ will be kept. To make sure that each element within the \ac{simd} unit is unsigned and minimize the numerical value of delta values within a group at the same time, we assign a group-wise y-axis intercept $n_j$: 
\begin{equation}
n_j = \min S_j
\end{equation}
which we can remove from the per-element values
\begin{equation}
S_j' = \lbrace \Delta c_i - n_j, \Delta c_{i+1} - n_j, \dots, \Delta c_{i+g-1} - n_j\rbrace
\end{equation}
At runtime, $n_j$ now describes the base pointer for the group. The offset and base pointer of an element $c_{i,j}$ in lane $i$ of \ac{simd} group $j$ can be reconstructed as 
\begin{equation}
c_{i,j} = \underbrace{i \cdot\left\lfloor k_d/k_s\right\rceil  + \Delta c_{i,j}}_{\text{scatter/gather offset}} + \underbrace{n_j}_{\text{Base Pointer}}
\end{equation}
Subtracting the minimum from $S_j$ brings all delta elements within $S_j'$ into the range $\left[0,\max S_j -\min S_j \right]$, ensuring that only the minimal number of bits is required to encode them. To reduce the memory footprint for each group further, we apply the same linear decomposition that we did above for elements in groups to the base pointer delta values $n_j$. The aim is to not encode the average distance between two \emph{groups} explicitly as it is redundant and usually large. The average distance between two groups is the number of elements in the group multiplied by the average distance between them $ g \cdot \lfloor k_d/k_s \rceil$. The y-axis intercept can be omitted as long as matrix density is not highly irregular. To reconstruct $n_j$ with zero-based indexing, the calculation of $n_j$ can be expressed as
\begin{equation}
n_j = j \cdot \left\lfloor \frac{k_d}{k_s} \right\rceil \cdot g + \Delta n_j
\end{equation}
Alternatively, $n_j$ can be stored incrementally, similar to the encoding of column indices in Relative Indexing. 
\begin{equation}
n_j =
\begin{cases}
	\Delta n_{j} & \text{for } j=0\\
	n_{j-1} + \lfloor k_d/k_s\rceil \cdot g + \Delta n_{j} &\text{otherwise}
\end{cases}
\end{equation}

\begin{figure}[tb]
	\includegraphics[width=\linewidth]{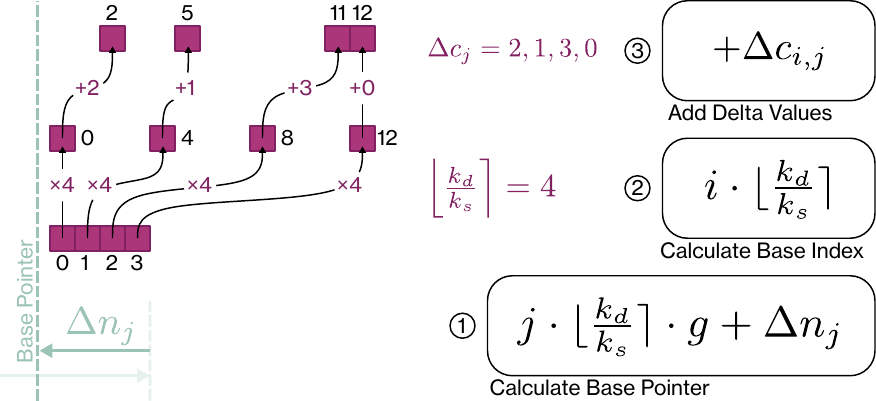}
	\caption{Runtime calculation of base pointer and scatter/gather offsets from Delta-Linear encoding for one \ac{simd} group}
	\label{fig:dcsr_decoding}
\end{figure}

In both cases, $\Delta n_{j}$ is the only part of the equation that needs to be put into memory. We find that the latter encoding produces a more even distribution of values with lower magnitude in our evaluation, although at the cost of introducing an interdependence between groups. The process of deriving a base pointer and per-element offsets from the delta representation is shown in~\cref{fig:dcsr_decoding}.\par
Some boundary conditions need to be ensured. The calculation of per-element offset values must not overflow the bounds given by the width of the \ac{simd} lane. In addition, $\Delta n_j$ also must not overflow its base type (an 8-bit signed integer in our implementation). We insert padding to ensure that all of these conditions hold. To do this, we employ a greedy algorithm that in each iteration inserts a zero value into the middle of the largest gap in the row until no overflow occurs.
\subsection{Dynamic Bitwidth Extension}
While \ac{dle} reduces the size of column index elements within a group, it can only guarantee the upper bound for these values that was achieved during the padding insertion. Simply encoding every column indices' delta value using the size of this upper bound would be too wasteful for many memory-constrained applications. Lower bitwidths that are not an even divisor of the machine word size (like five, six or seven bit) on the other hand cannot directly be represented in memory in a way that makes it easy to load them in parallel at runtime. Having the greatest element of a row determine the bitwidth for all other elements in a row also means that a single outlier value might lead to a lot of wasted memory when other elements would be representable with fewer bits.\par
\begin{figure}[tb]
	\includegraphics[width=\linewidth]{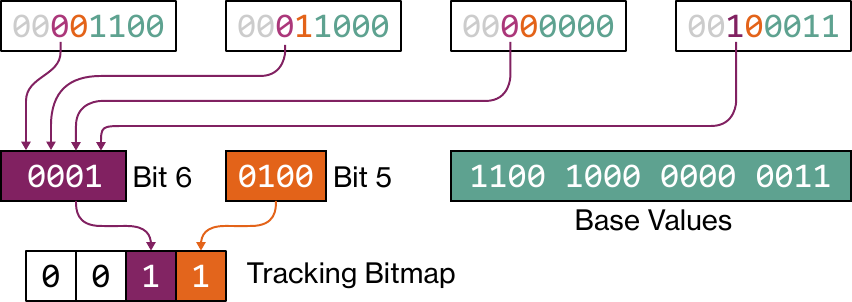}
	\caption{Decomposition of a \ac{simd} run that is encodable in six bits into 4-bit Base Values, two Extension Bitmasks and a Tracking Bitmap}
	\label{fig:decomposition}
\end{figure}

To address these issues, we decompose each group of delta elements into a base value with a fixed number of bits as well as several groups of \emph{extension bitmasks} as shown in~\cref{fig:decomposition}. Every bitmask contains one bit per \ac{simd} lane that marks whether or not the bit is set in the encoded value. Masks that would be all-zero are not created. The extension bits can dynamically be added to the base value at runtime if more bits than available in the base value are required. The decomposition is done separately for each \ac{simd} run. Through this, each \ac{simd} run can be encoded in the number of bits required for the largest element in this run, reducing the impact of rare outlier values to a single \ac{simd} group rather than an entire row. The size of the mask is determined by the number of \ac{simd} lanes which can be expected to be a power of two; the extension bitmasks do not cause issues with memory alignment because of this. Because the number of extension bitmasks is not fixed per group, we need to be able to infer how many of them need to be consumed for a given group at runtime. Each group maintains a \emph{tracking bitmap} for this purpose. This tracking bitmap has one bit for every bit position that might be extended. If the bit is set to one, there exists a bitmask for this position that needs to be consumed at runtime; if it is set to zero, there is no bitmask for this position.\par
\begin{figure}[tb]
	\includegraphics[width=\linewidth]{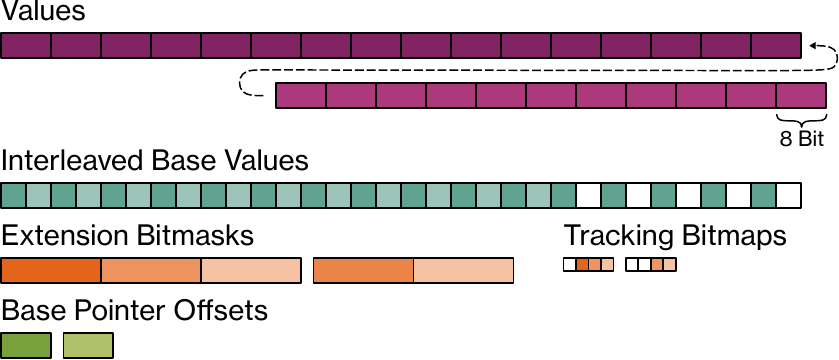}
	\caption{Memory representation of two consecutive \ac{simd} groups in \ac{dcsr}. Element sizes drawn to relative scale for ARM \ac{mve} implementation with 16 \ac{simd} lanes, 8-bit quantization and 4-bit base values. Interleaved storage of base values leads to partially unused index base value positions for the shorter second group.}
	\label{fig:memory_representation}
\end{figure}

For our implementation, we choose a base value size of four bits, with bits five, six and seven being extensible. A fixed-size base value is not strictly necessary for the algorithm, as any integer could also fully be decomposed into only a set of extension bitmasks. We introduce the base values because the re-composition at runtime incurs some overhead. It therefore makes sense to only dynamically add bit positions that have a high likelihood of not being present in a significant number of groups; for the lowest four bits, we find that this is not the case which is why they are always stored as-is. To make the 4-bit base values easily accessible in memory, we interleave two adjacent groups so that the base values of the first group occupy the upper nibble, while the base values of the second group occupy the lower nibble as shown in~\cref{fig:memory_representation}. The benefit of this scheme is that when executing a parallel load, each base value is loaded into the correct \ac{simd} lane. Accessing the values for either group can be done through a single parallel shift or bitwise-or operation. The bitmasks occupy a multiple of a byte and are stored contiguously without interleaving.\par
\begin{figure}[htb]
	\includegraphics[width=\linewidth]{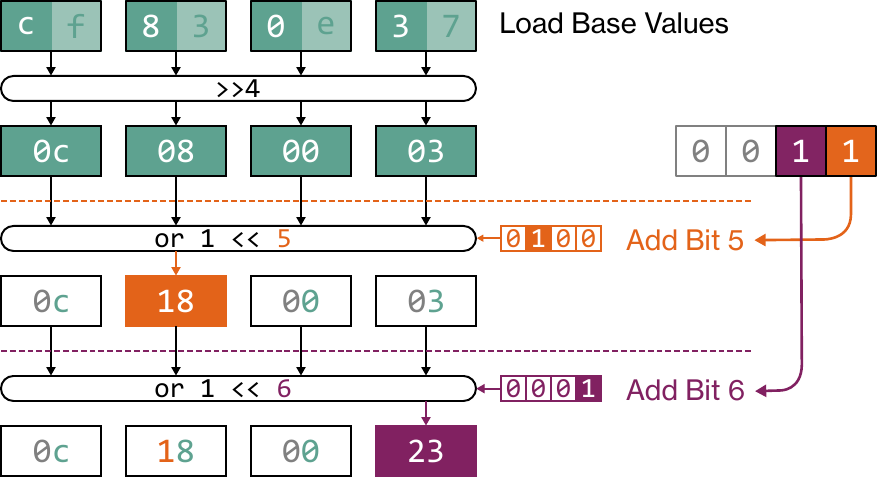}
	\caption{Parallel Reconstruction of a group delta values from groupwise interleaved base values, tracking bitmap and extension bitmasks}
	\label{fig:recomposition}
\end{figure}

We show the inverse process of combining tracking bitmap, extension bitmasks and base values into delta values in~\cref{fig:recomposition}. When recomposing the original value at runtime, we iterate over all extensible bit positions. If the corresponding bit is not set in the tracking bitmap the iteration is complete and we continue with the next position. If the bit is set we load the next bitmask from memory and advance the mask pointer. We then do a parallel bitwise-or operation with the current bit position. The mask is used to apply this operation only to the lanes that need to be extended in this group. The binary lane masking feature that is essential for this is present in ARM \ac{mve} as well as the RISC-V Vector Extension.\par
\subsection{Row Buffering}
\label{ssec:buffering}
While the extraction process of indices is lightweight, it is beneficial to reuse intermediate results when possible. In \ac{pwc} layers, every row of the sparse matrix is multiplied with every column of the activation matrix, allowing for repeated use of the same row of weights once decoded. Keeping element indices in memory once they are generated is therefore an attractive option to reduce the overhead of the extraction process further. Buffering can be carried out in one of two ways.

\paragraph{Value Buffering}
In \ac{vb} mode, the indices for the current row are generated from the base values, tracking bitmaps and bitmasks. Furthermore, the values for the current row are loaded from memory. The indices are then used to scatter the matrix values into the appropriate positions of the zero-initialized row buffer. The row buffer is now the reconstruction of the same row as it would appear in a dense matrix and can be multiplied with the input activation using dense kernels. In this mode, the \ac{dcsr} kernel is a simple compression/decompression scheme that cannot benefit from the reduced number of effective computations in a sparse matrix by skipping multiplications with zero. In practice, this can be preferable because gather memory accesses are significantly slower than dense and aligned memory accesses on most hardware. The reduced number of operations therefore might only outweigh the slow gather instructions for high sparsities.
\paragraph{Index Buffering}
When executing in \ac{ib} mode, the indices themselves are recomposed from their \ac{dbe} representation and written to the row buffer. The base pointer steps are also computed and buffered. The inference kernel then loads the pre-computed base pointers and gather offsets from memory and uses them to load the corresponding activation values using a gather memory access. In this mode, only operations with non-zero weights are carried out at the cost of potentially slower memory accesses.\par
The size for the row buffer can be upper-bounded to the size of one matrix row ahead of time. \ac{ib} requires some additional memory for base pointer steps which is upper-bounded by the maximum number of groups in a row. 
The equilibrium point between the two modes depends on the matrix characteristics as well as the target hardware. For \ac{fc} layers, a similar reuse of weights is not possible since each element in the weight matrix is only used exactly once. The \ac{fc} kernel will therefore always use extracted values directly to generate a gather instruction into the activation memory.

\section{Results and Discussion}
\subsection{Setup}
We use the set of \acp{dscnn} for keyword spotting from~\cite{zhang2018hello} as detailed in~\cref{tab:architectures} as a reference. 
\begin{table*}[tb]
	\caption{Reference Network Architectures}
	\footnotesize
	\begin{tabularx}{\textwidth}{llccc}
		\toprule
		Layer & Parameters & S & M & L \\
		\midrule
\phantom{$\rightarrow $} $1 \times$ 2D Convolution  & Channels, Kernel Size, Stride & 64, (10,4), (2,2) & 172, (10,4), (2,1) & 276, (10,4), (2,1) \\
$\rightarrow n \times$ 2D \ac{dsc} & Number of Blocks $(n)$, Channels, Kernel Size, Stride & 4, 64, (3,3), (1,1) & 4, 172, (3,3), (1,1) & 5, 276, (3,3), (1,1) \\
$\rightarrow 1 \times$ 2D AveragePooling & Size & (2,2) & (2,2) & (2,2) \\
$\rightarrow 1 \times$ Fully-Connected & Size & (12, 1536) & (12, 10320) & (12, 16560) \\
		\bottomrule
	\end{tabularx}
	\label{tab:architectures}
\end{table*}
For each network architecture, we train a dense base model that is subsequently pruned to increasing levels of sparsity. We deviate from the original architecture by adding Dropout~\cite{srivastava2014dropout} after the input convolution as well as each depthwise separable convolution block to address the overfitting seen in the original publication. We also omit an increased stride for the first depthwise separable block in the larger architectures. All base models are trained for 50 epochs on the Google Speech Commands v2 dataset~\cite{warden2018speech}. The models are trained using an Adam Optimizer with a linear learning rate decay from an initial rate of $5 \cdot 10^{-3}$ to the final learning rate of $1 \cdot 10^{-4}$.\par
\begin{figure}[tb]
	\centering
	\includegraphics[width=\linewidth]{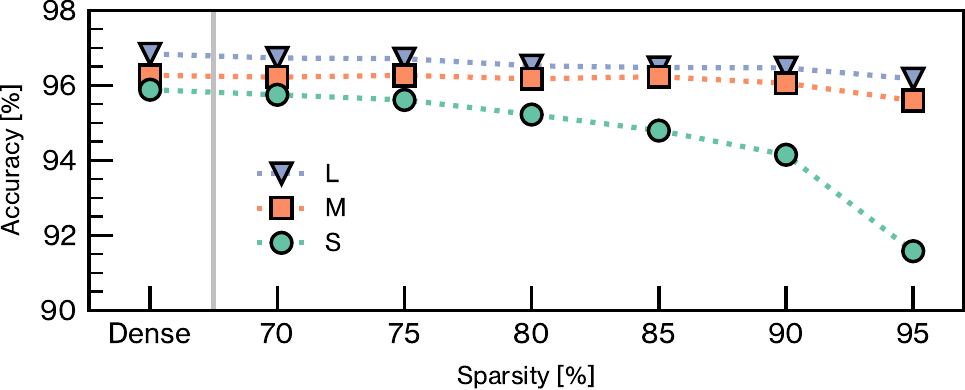}
	\caption{Degradation of accuracy through pruning of \ac{pwc} and \ac{fc} layers per DS-CNN architecture. Accuracies reported for floating-point models.}
	\label{fig:degradation}
\end{figure}

The base network is pruned to sparsities between 70\% and 95\% in increments of 5\% (the term \emph{sparsity} refers to the percentage of elements that are set to zero here). We prune only the weights of the \ac{pwc} and \ac{fc} layers in the network because they account for the majority of the network's memory footprint. The input convolution as well as the depthwise convolutions and all biases remain dense. The low-magnitude pruning layer from the Tensorflow~\cite{abadi2016tensorflow} Model Optimization Toolkit is used to prune each network over 40 epochs followed by 40 epochs of retraining. 
We show the degradation of accuracy for each model in~\cref{fig:degradation}. 
By targeting only layers with high parameter counts, we can achieve noticeable reductions in the number of active parameters for these layers without a significant impact on the model's accuracy.\par
All training scripts, conversion tools to \ac{dcsr} and \ac{tflite} micro~\cite{david2020tensorflow} inference kernels as well as the dense and sparse reference models are made publicly available~\cite{trommer21}.\par
The results in~\cref{tab:results} show that our base models achieve noticeably higher accuracies than the ones published in~\cite{zhang2018hello}. We attribute this effect primarily to the fact that our models were trained on the enlarged version 2 of the dataset with 105k samples---significantly more than the 65k samples that were available in the first version of the dataset.\par
For a detailed investigation, we pick the sparsest models that are within 1\% of the dense model's accuracy after pruning and quantization. This corresponds to 80\% sparsity for the \ac{s} architecture and 90\% sparsity for the \ac{m} and \ac{l} architectures.
\begin{table*}[tb]
	\caption{Comparison of execution cycles, accuracy and memory footprints for dense and sparse Models}
	\footnotesize
	
\begin{tabularx}{\textwidth}{YRRRRYYYYYYYYYY}

\toprule
{} & \multicolumn{4}{c}{Cycles $\cdot 10^{-6}$} & \multicolumn{1}{c}{Sparsity} & \multicolumn{2}{c}{Accuracy [\%]} &\multicolumn{7}{c}{Memory Footprint [KiB]}\\
\cmidrule(lr){2-5}
\cmidrule(lr){6-6}
\cmidrule(lr){7-8}
\cmidrule(lr){9-15}
Model &  dCSR (VB) &  Relative \newline Indexing & CMSIS \newline (opt.) & CMSIS & {}  & Dense, \newline FP32 &  Pruned,\newline Int8 & Total\newline(Dense) & PWC+FC (Dense) & PWC+FC (Pruned) & Metadata (dCSR) &  Metadata (RI) & Padding (dCSR) & Padding (RI)\\
\midrule
L &        60.97 &      64.27 &               65.32 &        104.64 &        90\% &                   96.84 &                    96.08 &      617.01 &                     566.02 &                       54.27 &            52.40 &          36.61 &            1.87 &         13.96 \\
M &        28.69 &      29.88 &               30.08 &         42.94 &        90\% &                   96.27 &                    95.32 &      265.45 &                     236.50 &                       23.61 &            23.90 &          15.50 &            0.90 &          5.17 \\
S &         5.54 &       5.71 &                5.68 &          6.94 &        80\% &                   95.88 &                    94.91 &       42.56 &                      34.00 &                        6.79 &             5.94 &           4.01 &            0.00 &          0.24 \\
\bottomrule
\end{tabularx}
	\label{tab:results}
\end{table*}
We provide numbers for end-to-end execution cycle counts for the most relevant algorithms in~\cref{tab:results}. These numbers reflect the performance on the surveyed models, but can not offer insights into the performance for different layer types. The achieved compression ratios depend on matrix characteristics and are similar across different layer types. The effects on throughput for \ac{pwc} and \ac{fc} layers, however, are not, due to their different memory access patterns. For this reason, we provide a separate evaluation of the inference speed for both target layer types in \cref{ssec:spmm} and \cref{ssec:spmv}.

\subsection{Compression}
To assess the memory footprint of \ac{dcsr}, we compare its overall memory consumption with that of other established techniques. We include \ac{csr} and \ac{bcsr} for reference, even though they are not optimized for low overhead. While \ac{csr} is the de facto standard for scientific computation on sparse matrices, a variant of \ac{bcsr} is used to encode sparse tensors in \ac{tflite} (\emph{not} \ac{tflite} micro, where there is currently no support for sparse tensors). For \ac{csr} and \ac{bcsr}, we assume 16-bit values for the row pointer and column index arrays because it is the smallest memory-aligned integer size that can encode all layers in all surveyed models. For \ac{bcsr}, we further assume a block size of $(2,2)$. Reported compression ratios are after quantization of weights to eight bit. Memory footprints are based on the sizes of a model's weights and biases and exclude overhead from the \ac{tflite} Flatbuffer data structure.\par
\begin{figure}[tb]
	\centering
	\includegraphics[width=\linewidth]{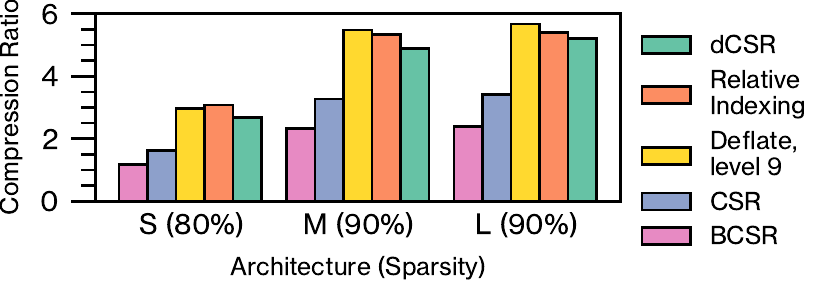}
	\caption{Comparison of total compression ratio for target layers per architecture}
	\label{fig:compression_ratios}
\end{figure}

The low compression ratios for \ac{csr} and \ac{bcsr} in~\cref{fig:compression_ratios} indicate that both formats produce significant overhead. Relative Indexing is superior to general-purpose compression for the small model and achieves similar results for the medium and large architecture with 90\% sparsity. \ac{dcsr} does not achieve the same compression ratios as Relative Indexing for the small model with lower sparsity. This gap closes increasingly for the two larger architectures with higher sparsity.\par
A breakdown of the contributions of metadata, padding elements and active weights to the overall memory footprint of the compressed sparse matrix is given in~\cref{tab:results}. Relative Indexing produces lower amounts of metadata per non-zero element, but does so at the cost of inserting a high amount of padding into the values arrays. For the small architecture, the padding overhead of Relative Indexing is 3.6\% of the values array, while the medium and large models require the insertion of 21.9\% and 25.7\% additional zero values. \ac{dcsr} also inserts padding, but its greedy search results in a much lower amount. The inserted padding of \ac{dcsr} totals 0\% (S), 3.8\% (M) and 3.4\% (L) additional elements. The larger overhead for \ac{dcsr} altogether---despite the higher amount of padding inserted for Relative Indexing---can be explained by two things: smaller matrices incur more overhead in \ac{dcsr} because base indices need to be stored, even for partially filled \ac{simd} groups like we illustrate in~\cref{fig:memory_representation}. Handling metadata like tracking bitmaps and base pointer delta values facilitate fast execution at runtime, but increase the overhead further. Also, \ac{dcsr} can only limit the effect of outlier values to a group, not individual elements. A single outlier in a \ac{simd} group still causes the creation of extension bitmasks for all the other elements in the group. A strength of \ac{dcsr} is that it handles matrices with regions of high sparsity better through its flexible encoding of bitwidths, resulting in fewer padding elements being inserted for larger, more sparse matrices.
\subsection{Embedded SpMM}
\label{ssec:spmm}
\begin{figure}[tb]
	\centering
	\includegraphics[width=\linewidth]{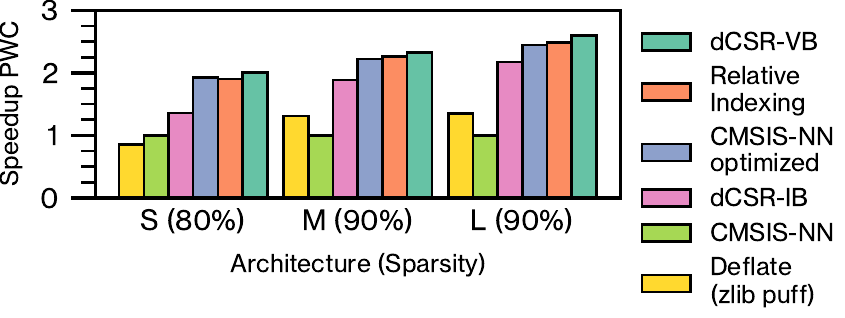}
	\caption{Total Speedup of \ac{pwc} operations over CMSIS-NN}
	\label{fig:conv_speedup}
\end{figure}

To analyze the overhead of our algorithm, we demonstrate the extraction and sparse inference as dedicated kernels within the \ac{tflite} micro framework. The compression of supported weight tensors inside a \ac{tflite} model takes place ahead of time as part of the model conversion. Extraction and inference is implemented in C using ARM \ac{mve} intrinsics. We report the cycle count as measured by the \ac{pmu} on the ARM MPS3 AN547 FPGA prototype of the ARM M55 processor. What makes the M55 highly relevant for \ac{nn} inference is its updated set of \ac{simd} instructions. While the \ac{dsp} instructions of previous generations of Cortex-M devices only support 16-bit \ac{mac} operations, \ac{mve} adds support for 8-bit \acp{mac}---a fundamental operation in quantized \ac{nn} inference. In addition, \ac{mve} instructions operate on 128-bit words, four times wider than ARM DSP instructions. This results in much higher throughput for highly parallel applications like \acp{nn}.
Not all models fit the system's \ac{bram}. To keep the numbers comparable, all models are executed out of the system's DDR4 RAM at runtime. We compare our implementation to a scalar calculation of Relative Indexing offsets and dense ARM \ac{mve} kernels from the CMSIS-NN framework~\cite{lai2018cmsis}.  We also compare the results to the \emph{puff} inflate algorithm from the zlib compression library~\cite{zlib1996}. \emph{puff} is a minimal reference implementation that supports the full Deflate algorithm~\cite{deflate1996} in a vastly reduced code size at the cost of lower throughput. We prefer it over the full implementation because the reduced code size makes it a more appropriate benchmark in the context of embedded systems.
We first analyze the networks' \ac{pwc} operations which are instances of the \ac{spmm}. To make the numbers comparable between different models, we normalize execution cycle counts to those achieved by CMSIS-NN. In our results in~\cref{fig:conv_speedup}, we find that our value-buffered implementation achieves a speedup of more than two over the \ac{pwc} kernel in CMSIS-NN. This is unexpected since the amount of arithmetic operations for both is identical with additional operations for the unpacking of the value-buffered \ac{dcsr}. The reason lies in the iteration order of the CMSIS-NN kernel: it iterates over filter weights in the inner loop which need to be loaded from the slower DDR4 RAM repeatedly as a consequence. Our implementation in contrast iterates over all activations for a buffered row of filter values. Both reside in the faster \ac{bram}. We also include an adapted version of the CMSIS-NN kernel that adopts an iteration order and buffering scheme similar to ours for the dense matrix in the evaluation (CMSIS-NN optimized). Even with this optimization in place, \ac{dcsr} outperforms CMSIS-NN and Relative Indexing with scalar unpacking. The difference in execution speeds underlines the vastly decreased number of memory accesses in \ac{dcsr}.\par
We find that our index-buffered implementation does not reach the equilibrium point at which the slow gather memory accesses produce lower overhead compared to a dense kernel for the evaluated hardware and sparsities. The speedup achieved by value-buffered \ac{dcsr} over the optimized CMSIS-NN kernel despite the increased number of arithmetic instructions is largely due to the reduced pressure on the system's memory interface.\par
\begin{figure}[tb]
	\centering
	\includegraphics[width=\linewidth]{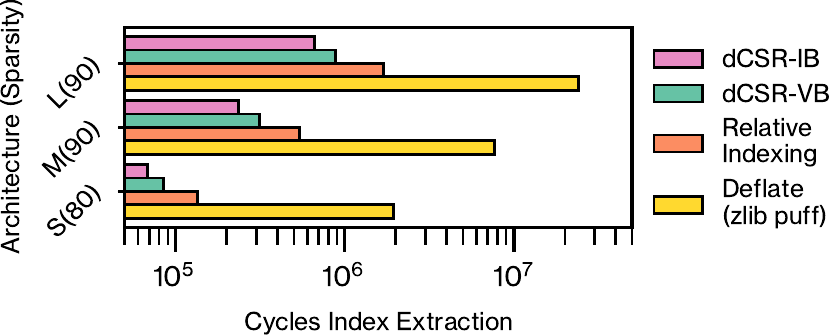}
	\caption{Comparison of index extraction cycle counts for \ac{pwc} layers}
	\label{fig:extraction}
\end{figure}

A closer look at the time spent in the extraction stage in~\cref{fig:extraction} shows that both implementations of \ac{dcsr} only require around 50\% of the cycles needed by Relative Indexing to restore matrix indices from their representation in memory, with the index-buffered variant requiring 20\% to 25\% fewer cycles when compared to the value-buffered version due to not having to load weight values from memory. Lossless compression/decompression based on the general-purpose Deflate algorithm incurs significant overhead when compared to dedicated sparse matrix storage solutions. This suggests that general-purpose decompression as part of the inference process is not competitive with dedicated sparse matrix encoding in terms of throughput for software-only implementations.
\subsection{Embedded SpMV}
\label{ssec:spmv}
\begin{figure}[tb]
	\centering
	\includegraphics[width=\linewidth]{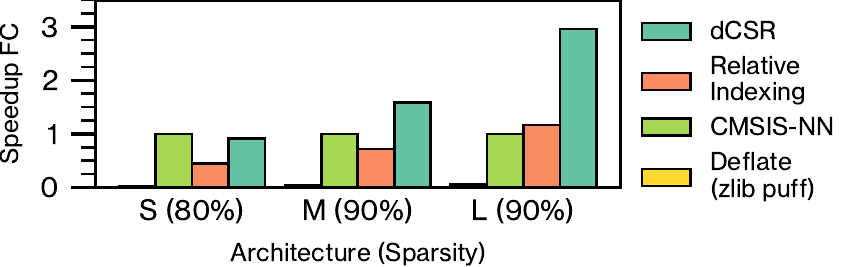}
	\caption{Speedup of \ac{fc} layer inference over CMSIS-NN}
	\label{fig:fc_speedup}
\end{figure}

Another important building block of \acp{nn} is the \ac{spmv} operation. Different from \ac{spmm}, this workload is memory-bound~\cite{williams2007}. Each \ac{dscnn} architecture implements an \ac{fc} layer as the last layer of the network, which is an instance of the \ac{spmv}. For the Relative Indexing implementation, we extract the indices for one \ac{simd} run in scalar code, transfer them to the \ac{simd} unit and use them to generate a gather memory access into the activations. We compare this with direct \ac{simd} decompression in \ac{dcsr} as well as the dense CMSIS-NN kernel in~\cref{fig:fc_speedup}. An evaluation of different buffering schemes is not applicable for \ac{fc} layers (see \cref{ssec:buffering}). The \ac{dcsr} implementation is slightly slower than CMSIS-NN for the smallest matrix with lower sparsity, but shows a noticeable speedup for the two larger \ac{fc} layers with higher sparsity. In SpMV, a  balance of a fast unpacking process in combination with high compression of weights is essential since slow memory accesses cannot be amortized. In contrast to Relative Indexing, \ac{dcsr} has two other benefits that help in achieving a high throughput. The index offset values already reside in the \ac{simd} unit after calculation; in Relative Indexing, they need to be calculated in the scalar domain and then transferred to the \ac{simd} unit. Secondly, the larger amount of padding for Relative Indexing further reduces throughput.

\subsection{Case Study: Sparse MobileNetV2}
We apply a similar strategy as above to the MobileNetV2~\cite{sandler2018mobilenet} architecture. We initially tune a dense network that was pre-trained on the Imagenet~\cite{deng2009imagenet} dataset for classification on the CIFAR-10 task~\cite{krizhevsky2009learning} with images scaled up to $96 \text{px} \times 96 \text{px}$ as a \ac{bl}. We prune only the last \ac{pwc} and \ac{fc} layers in the network (from the 12\textsuperscript{th} and 11\textsuperscript{th} inverted residual block onward for 80\% and 93\% sparse networks respectively). The weights of these layers account for the 77.2\% and 82.0\% of the model's memory footprint after 8-bit quantization, so pruning them will matter most. In addition, we find that pruning early layers with low parameter counts disproportionately reduces the model's final accuracy. We compare the memory footprint after pruning and encoding of the sparse layers with \ac{dcsr} to those of dense MobileNetV2 models with a reduced width multiplier in \cref{tab:mnetv2}. Our evaluation confirms that large sparse networks offer a better accuracy with lower memory consumption compared to their small dense counterparts. This observation is consistent with findings in previous work~\cite{zhu2017prune}. A drawback of large sparse architectures is that they cannot profit from reduced arithmetic operations to the same degree that small dense networks can. Sparsity is a vital tool, however, when maximizing accuracy in a given memory envelope. \par
\begin{figure}[tb]
	\centering
	\includegraphics[width=\linewidth]{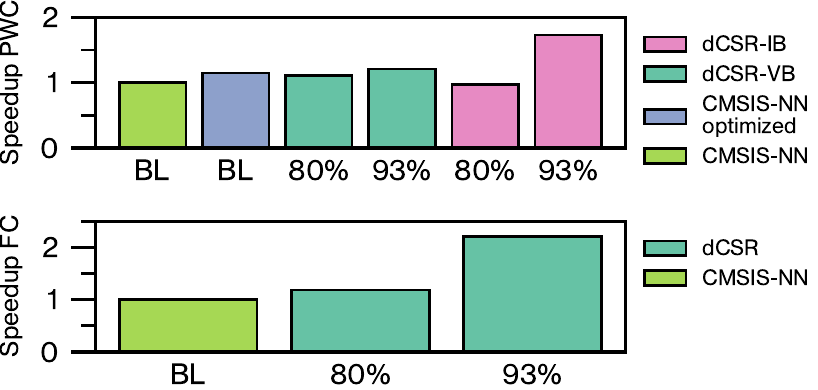}
	\caption{Speedup for target layers on sparse MobileNetV2 over CMSIS-NN}
	\label{fig:mnet_speedup_pwc}
\end{figure}

We compare the speedup in total cycles for both target layer types in~\cref{fig:mnet_speedup_pwc}. For the 80\% sparse MobileNetV2, the performance of \ac{dcsr} for \ac{pwc} is slightly below that of the optimized CMSIS. This is because some of the weight matrices are of irregular shape with many more rows than columns, which incurs the per-row overhead in \ac{dcsr} more often. Because of the increased sparsity in the 93\% sparse model, index-buffered \ac{dcsr} now yields a significant speedup over other implementations through omitting multiplications with zero. The results for the networks' \ac{fc} layer corroborate our findings from the keyword spotting application in that high sparsities result in increased throughput over dense inference for \ac{dcsr}-encoded matrices.
\begin{table}[tb]
	\caption{Comparison of large sparse and small dense MobileNetV2 on CIFAR-10}
	\footnotesize
	\begin{tabularx}{\linewidth}{lrYYr}
	\toprule
	{} & Sparsity & Width Multiplier & Accuracy [\%] & Size [KiB] \\
	\midrule
	(BL) & 0\% & 1.0 & 94.44 & 2220.3\\ 
	{} & 0\% & 0.75 & 93.89 & 1364.9\\
	{} & 0\% & 0.5 & 92.19 & 704.9\\ \midrule
	{} & 80\% & 1.0 & 94.03 & 1127.1\\
	{} & 93\% & 1.0 & 92.69 & 695.7 \\

	\bottomrule
\end{tabularx}
\label{tab:mnetv2}
\end{table}

\section{Conclusion}
In this work, we introduced \ac{dcsr}, a novel sparse matrix representation that helps materialize the benefits of pruning in embedded hardware with \ac{simd} capabilities. We showed that even small \acp{nn} might still be over-parameterized and can be pruned to significant sparsities at a marginal loss in accuracy. For a typical embedded \ac{nn} application, \ac{dcsr} achieves a reduction in size of 62.6\% to 80.8\% for the targeted layers and 50.0\% to 74.1\% for the entire network after quantization. This is complemented by an increase in end-to-end throughput on our target hardware. The speedup is achieved through reduced memory transactions combined with a parallel extraction of indices from their representation in storage through \ac{simd} instructions. Especially for \ac{spmv} operations in \acp{nn}, which are memory-bound and often account for a large portion of the network weights, our method promises a significant reduction both in size and inference time. Looking at the current trajectory of embedded systems, it seems likely that they will follow the path of desktop systems with compute capacity growing at a more rapid pace than memory, both in speed and availability. The \ac{dcsr} method shows a way of closing this coming gap for \acp{nn} by harnessing the increasing computational potential to reduce memory consumption.

\IEEEtriggeratref{16}
\newpage
\bibliographystyle{IEEEtran}
\bibliography{IEEEabrv, bibliography}

\end{document}